\documentclass[twocolumn]{revtex4}

\usepackage{graphicx}
\usepackage{amsmath,amsthm,amssymb,bm,amsfonts}

\def\qbar{{\overline{q}}}
\def\Phat{{P}}

\def\fprime{{f^\prime}}

\def\Kcal{{\cal{K}}}

\newcommand{\xzt}{\bigg(\frac{x}{z},t\bigg)}

\begin{document}

\title{\bf How to impose initial conditions for QCD evolution \\of double parton distributions?}

\author{Krzysztof Golec-Biernat}\email{golec@ifj.edu.pl}
\affiliation{Institute of Nuclear Physics Polish Academy of Sciences, 31-342 Cracow, Poland}
\affiliation{Faculty of Mathematics and Natural Sciences, University of Rzesz\'ow, 35-959 Rzesz\'ow, Poland}

\author{Emilia Lewandowska}\email{emilia.lewandowska@ifj.edu.pl}
\affiliation{Institute of Nuclear Physics Polish Academy of Sciences, 31-342 Cracow, Poland}

\begin{abstract}
Double parton distribution functions are used in the QCD description of double parton scattering. 
The double parton distributions evolve with hard scales through QCD evolution equations which obey nontrivial 
momentum and valence quark number sum rules. 
We describe  an attempt to construct initial conditions for the evolution equations which exactly fulfill 
these  sum rules and discuss its shortcomings.
We also discuss the factorization of the double parton distributions  into a product of two single parton distribution functions
at small values of the parton momentum fractions.

\end{abstract}


\maketitle

\section{Introduction}

In high-energy  hadron scattering,
the final state particles could be produced from two hard interactions in one collision.
This process, called the  double parton scattering (DPS), 
is viewed as two hard interactions of two pairs of partons from the scattering hadrons.
The DPS is the simplest process in the analysis of multiparton interactions, studied for many years 
from both the theoretical \cite{Kirschner:1979im,Shelest:1982dg,Zinovev:1982be,Ellis:1982cd,Bukhvostov:1985rn,
Snigirev:2003cq,Korotkikh:2004bz,Gaunt:2009re,Blok:2010ge,Ceccopieri:2010kg,Diehl:2011tt,
Gaunt:2011xd,Ryskin:2011kk,Bartels:2011qi,Blok:2011bu,Diehl:2011yj,Manohar:2012jr,Ryskin:2012qx,Gaunt:2012dd,Snigirev:2014eua} 
and phenomenological sides
\cite{DelFabbro:1999tf,Kulesza:1999zh,DelFabbro:2002pw,Cattaruzza:2005nu,Berger:2009cm,Gaunt:2010pi,Kom:2011bd,Berger:2011ep,Kom:2011nu, Bartalini:2011jp,d'Enterria:2012qx,d'Enterria:2013ck}. 
The experimental evi\-dence of the DPS has been presented 
in Refs.~\cite{Akesson:1986iv,Abe:1997bp,Abe:1997xk,Abazov:2009gc,Aad:2013bjm,Chatrchyan:2013xxa,Aad:2014rua}.

The DPS processes allow one to gain information on parton correlations by  measuring the DPS cross section 
in high energy scattering of two hadrons,
$h_1$ and $h_2$. In the collinear approximation, the inclusive DPS cross section is given 
in the form \cite{Diehl:2011yj,Ryskin:2011kk}:
\begin{eqnarray}\nonumber
\label{cross1}
\sigma_{AB}^{h_1h_2} \!\!&=&\!\!\frac{N}{2}
\sum_{f_1f_2f_1^\prime f_2^\prime}\int dx_1dx_2\,dx^\prime_1dx^\prime_2\, \frac{d^2{\bf q}}{(2\pi)^2}
\\\nonumber
&\times&\!\!D_{f_1f_2}^{h_1} (x_1,x_2,Q_1,Q_2,{\bf q})\, {\hat{\sigma}}^A_{f_1f_1^\prime}(x_1,x^\prime_1,Q_1)
\\\nonumber
\\
&\times&\!\!{\hat{\sigma}}^B_{f_2f_2^\prime}(x_2,x^\prime_2,Q_2)\, D_{f_1^\prime f_2^\prime}^{h_2}(x^\prime_1,x^\prime_2,Q_1,Q_2,-{\bf q}),~~~~
\end{eqnarray}
where $A$ and $B$ denote the two hard parton processes and $N$ is a symmetry factor, equal to 1 for $A=B$
and 2 otherwise.

In the above,  $D_{f_1f_2}^{h_{1,2}}(x_1,x_2,Q_1,Q,_2,{\bf q})$ are double parton distribution functions (DPDFs) of  hadrons $h_{1,2}$, 
which depend on the two parton flavors $f_{1,2}$, parton  momentum fractions $x_{1,2}$,  
two hard scales $Q_{1,2}$ involved  in the DPS  and an additional transverse momentum ${\bf q}$. 
The presence of the latter momentum is related to the loop structure of the exchanged
four partons in the forward scattering amplitude   which ultimately enters into the definition of the DPDFs 
\cite{Diehl:2011yj}. The importance of this variable for the DPS cross section computations has been
discussed at length  in Ref.~\cite{Ryskin:2011kk}.
The  longitudinal momentum fractions  obey the condition
\begin{eqnarray}
\label{eq:limit}
0< x_1+x_2 \le 1\,,
\end{eqnarray}
which says that the sum of parton longitudinal momenta cannot exceed the total proton momentum (taken for definiteness from now on).
This is the basic parton correlation which has to be taken into account. For more advanced aspects of parton correlations, see  Ref.~\cite{Manohar:2012jr}.

The analysis of the DPS is crucial for  a better understanding of background for many important processes measured at 
the experiments at Tevatron and the LHC,  e.g. for the Higgs boson production \cite{Krasny:2013aca}, as well as for  a better description of multiparton
interactions, needed for example for modeling of the underlying event;
see Ref.~\cite{Bartalini:2011jp} for a comprehensive review of these issues. 
Thus, it is very important to use a rigorous approach based on QCD evolution equations 
for the DPDFs.  These equations are known in the leading logarithmic approximation \cite{Kirschner:1979im,Shelest:1982dg,Zinovev:1982be, Snigirev:2003cq,Korotkikh:2004bz}.
They conserve new sum rules \cite{Gaunt:2009re} which relate the double and single parton distribution functions at any evolution scale.
In this presentation we address a problem of specifying  initial conditions for the evolution equations
of the DPDFs which  exactly obey the new sum rules.

The paper is organized as follows. In Secs.~II and III we briefly describe evolution equations for parton distributions.
In Sec.~IV the new sum rules are presented, while in Sec.~V the most popular initial conditions are described.
In Sec.~VI we discuss a problem with them and describe an attempt to solve it. In Sec.~VII factorization of the DPDFs
into a product of two single parton distribution functions (SPDFs) is discussed.  


\section{Evolution equations for SPDFs}

To set the notation, let us recapitulate the QCD evolution equations in the collinear approximation for  SPDFs, 
${D_f(x,Q)}$, which are used in  the description of the single parton scattering. 
The general form of these equations is given by
\begin{eqnarray}
\label{eq:onepdfeq}
\partial_{t}D_{f}(x,t)=\sum_{f^\prime}\int^{1}_{0}du\, {\cal{K}}_{ff^\prime}(x,u,t)\,D_{f^\prime}(u,t)\,,
\end{eqnarray}
where the evolution parameter $t=\ln(Q^2/Q_0^2)$ and the parton momentum fraction $x$ obey the condition
$0 < x \le 1$. 
The integral kernels, ${\cal{K}}_{ff^\prime}(x,u,t)$,  describe the real and virtual parton emissions
\begin{eqnarray}
\label{eq:num4}
{\cal{K}}_{ff^\prime}(x,u,t)={\cal{K}}_{ff^\prime}^R(x,u,t) - \delta(u-x)\,\delta_{ff^\prime}\,{\cal{K}}_{f}^V(x,t)\,.
\end{eqnarray}
The real emission kernel ${\cal{K}}_{f\fprime}^R(x,u,t)$ 
corresponds to the parton transition $(\fprime,u)\to (f,x)$, where the momentum fraction $u>x$,
and is given by 
\begin{eqnarray}
\label{eq:kreal}
{\cal{K}}^R_{ff'}(x,u,t)=\frac{1}{u} P_{ff'}(\frac{x}{u},t)\,\theta(u-x)\,.
\end{eqnarray}
The virtual part, ${\cal{K}}_{f}^V(x,t)$, can be computed from the imposed momentum sum rule 
\begin{eqnarray}
\label{eq:smom}
\sum_f \int^1_0 dx\,xD_f(x,t)=1
\end{eqnarray}
where the normalization to unity means that partons carry the whole nucleon momentum. Thus we find
\begin{eqnarray}
\label{eq:num6}
x\,\Kcal_{f}^V(x,t)=\sum_{\fprime} \int\limits_0^1 du\, u\,  \Kcal_{\fprime f}^R(u,x,t)\,.
\end{eqnarray}
The functions ${P_{ff'}}$  in Eq.~(\ref{eq:kreal}) are  splitting functions computed perturbatively 
in QCD in powers of the strong coupling constant:
\begin{eqnarray}
P_{ff'}(z,t)=\frac{\alpha_s(t)}{2\pi}P^{(0)}_{ff'}(z)+\frac{\alpha^2_s(t)}{(2\pi)^2}P^{(1)}_{ff'}(z)+...\,.
\end{eqnarray}
The first term  on the rhs corresponds to the leading logarithmic approximation while the higher terms are computed in the next-to-leading approximations.
In this way, the  well known DGLAP evolution equations for SPDFs are obtained
\begin{eqnarray}\nonumber
\label{eq:num10}
\partial_t\, D_f(x,t) \!\!\!&=&\!\!\! \sum_{\fprime}\int\limits_x^1 \frac{dz}{z}\,
\Phat_{f\fprime}(z,t) D_{\fprime}\xzt
\\
&-&\!\!\! D_f(x,t)\sum_{\fprime}\int\limits_0^1 dz z\,\Phat_{\fprime f}(z,t)\,.
\end{eqnarray}
Note that the diagonal in flavors splitting functions, $\Phat_{ff}(z,t)$, have a  simple pole singularity 
 at $z=1$ which is removed by the virtual term [so called $(+)$ prescription].

\begin{figure*}[t]
\centering\includegraphics[width = 11cm]{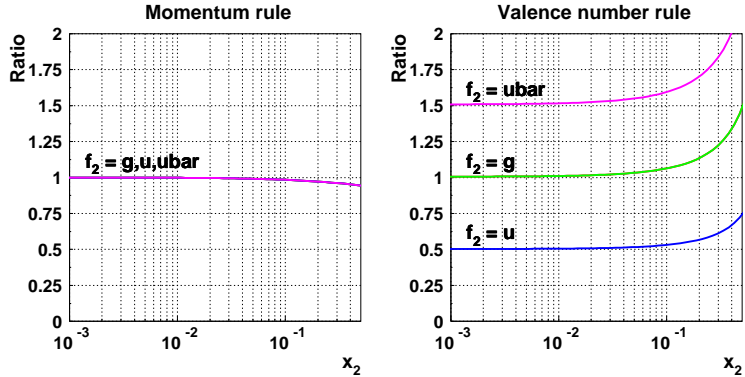}
\caption{Sum rules violation by the symmetric input (\ref{eq:gsinput}). The ratio should be equal to 1 if the sum rules are satisfied.}
\label{fig1}
\end{figure*}

\section{Evolution equations for DPDFs}

The evolution equations for the DPDFs  are only known  for ${\bf q}={\bf 0}$ in the
leading logarithmic approximation \cite{Kirschner:1979im,Shelest:1982dg,Zinovev:1982be,Snigirev:2003cq,Korotkikh:2004bz,Gaunt:2009re}. The first discussion of the next-to-leading corrections can be found in \cite{Ceccopieri:2010kg}.
We start from considering  two equal hard scales, $Q_1=Q_2\equiv Q$, and introduce
the following notation for the DPDFs in such a case
\begin{eqnarray}
D_{f_1f_2}(x_1,x_2,t)= D_{f_1f_2}(x_1,x_2,Q,Q,{\bf q}={\bf 0})\,.
\end{eqnarray}
A phenomenological discussion of the case ${\bf q}\ne {\bf 0}$ in the context of  evolution equations, discussed below, can be found in 
\cite{Ryskin:2011kk,Golec-Biernat:2014nsa}.

The QCD evolution equations take general form
\begin{eqnarray}\nonumber
\label{eq:twopdfeq}
\partial_t\!\!\!\!\!&&\!\!\!\!\!\! D_{f_1f_2}(x_1,x_2,t) 
 \\\nonumber
 &=&\!\!\sum_{f'}\int^{1-x_2}_{0} du \,{\cal{K}}_{f_1f'}(x_1,u,t) \, D_{f' f_2}(u,x_2,t)
\\\nonumber
&+&\!\!\sum_{f'}\int_{0}^{1-x_1}du\,{\cal{K}}_{f_2f'}(x_2,u,t) \,D_{f_1f'}(x_1,u,t)
\\\nonumber
\\
&+&\!\! \sum_{f'}\,{\cal{K}}_{f'\to f_1f_2}^R (x_1,x_2,t)\, D_{f'}(x_1+x_2,t),
\end{eqnarray}
where the integral kernels are given by Eq.(\ref{eq:num4}) with the real part (\ref{eq:kreal}) in the leading logarithmic
approximation and the virtual part found from Eq. (\ref{eq:num6}).
The two integrals in the above
describe the DGLAP evolution of a single parton with the second parton treated as
a spectator. This gives the upper integration limits resulting from condition (\ref{eq:limit}).
 
The third term needs special attention. It
describes the real emission splitting of a single parton into two partons which undergo two independent hard scatterings. This is why 
the SPDFs appear here and the evolution equations (\ref{eq:onepdfeq}) and (\ref{eq:twopdfeq}) form a coupled set
of equations which has to be solved simultaneously.
In the leading logarithmic approximation, there is only one parton flavor, $\fprime$, which leads to two 
parton flavors, $f_{1}$ and $f_2$. Thus, we have the following splittings: $q\to qg$,  $\qbar\to \qbar g$,  
$g\to q\qbar$, and $g\to gg$. In such  a case
\begin{eqnarray}\nonumber
{\cal{K}}_{f'\to f_1f_2}^R (x_1,x_2,t) = \frac{\alpha_s(t)}{2\pi}\,\frac{1}{x_1+x_2}\,P^{(0)}_{\fprime f_1}\!\!\left(\frac{x_1}{x_1+x_2}\right)
\\
\end{eqnarray}
where $P^{(0)}_{\fprime f_1}$ are splitting functions in the leading logarithmic approximation. It can  easily be checked that the splitting functions
$P^{(0)}_{\fprime f_2}(x_2/(x_1+x_2))$ can also be used in this case. Thus the rhs. of the evolution equations (\ref{eq:twopdfeq}) is invariant with respect to the parton interchange, $(f_1,x_1) \leftrightarrow (f_2,x_2)$.
If the initial conditions for them, specified at some initial scale $t_0$, are parton exchange symmetric,
\begin{eqnarray}
\label{eq:partonsym}
D_{f_1f_2}(x_1,x_2,t_0) = D_{f_2f_1}(x_2,x_1,t_0)\,,
\end{eqnarray}
the evolution  will preserve this symmetry for any value of $t$.

In the case when the two hard scales are significantly different, e.g. $Q_1 \ll Q_2$, the large logarithms $\ln(Q_2^2/Q_1^2)$ appear. They have to be  resummed which leads to the DGLAP evolution equation with respect to the second parton
\begin{eqnarray}\nonumber
\label{eq:num30a}
\partial_{t_2}\!\!\!\!&D&\!\!\!\!_{f_1f_2}(x_1,x_2,t_1,t_2)
\\
&=&\!\!\sum_{\fprime}  \int\limits_{0}^{1-x_1}{du}\,
\Kcal_{f_2\fprime}({x_2},{u},t_2)\,
D_{f_1\fprime}(x_1,u,t_1,t_2),~~~~
\end{eqnarray}
where $t_{1,2}=\ln(Q_{1,2}^2/Q_0^2)$. Thus the evolution has two steps, from equal initial scales $(t_0,t_0)$ 
to the equal final scales $(t_1,t_1)$,
according to Eq.~(\ref{eq:twopdfeq}), and then to the scales $(t_1,t_2)$, according to Eq.~(\ref{eq:num30a}). 
However, we do not discuss  such a  case in our analysis, concentrating only on the first step of the evolution.
We also refrain from discussing the impact parameter representation of the DPDFs and corresponding evolution equations, sending the reader to Ref.~\cite{Diehl:2011yj}.

\section{Sum rules for DPDFs}

The DGLAP evolution equations (\ref{eq:onepdfeq}) obey
the momentum sum rule (\ref{eq:smom}),
while the evolution equations (\ref{eq:twopdfeq}) preserve
a new  momentum sum rule: 
\begin{eqnarray}
\label{eq:momrule0}
\sum_{f_1}\int_{0}^{1-x_2}dx_1x_1\frac{D_{f_1f_2}(x_1,x_2,t)}{D_{f_2}(x_2,t)}=1-x_2\,.
\end{eqnarray}
This relation can be understood by treating
the ratio of the parton distributions under the integral as the conditional probability to find parton $f_1$ with  
the momentum fraction ${x_1}$, while the second  parton characteristics, $x_2$ and $f_2$, are  fixed.
 In such a  the total momentum fraction carried
by partons $f_1$ equals $(1-x_2)$.
In this way, the momentum sum rule (\ref{eq:momrule0}) relates the  double and single parton distribution functions for any value of $t$:
\begin{eqnarray}
\label{eq:momrule}
\sum_{f_1}\int_{0}^{1-x_2}\!\!\!dx_1x_1D_{f_1f_2}(x_1,x_2,t)\!=\!(1-x_2)D_{f_2}(x_2,t).~~
\end{eqnarray}

The valence quark  number sum rule for the SPDFs has the well-known form
\begin{eqnarray}
\int_0^{1}dx\left\{D_{q_i}(x,t)-D_{\bar{q_i}}(x,t)\right\}=N_{i}\,,
\end{eqnarray}
where $N_i$ is the number of valence quarks $q_i$.
For the DPDFs, the  analogous sum rule depends  on the flavor of the second parton $f_2$ (see 
Refs.~\cite{Gaunt:2009re,Gaunt:thesis}
for more details):
\begin{eqnarray}\nonumber 
\int_0^{1-x_2}\!\!\!\!\!\!&&\!\!\!\!\!\!dx_1\!\left\{D_{q_if_2}(x_1,x_2,t)-D_{\bar{q_i}f_2}(x_1,x_2,t)\right\} 
\\
&=&
\left\{
\begin{array}{ll}
  N_{i}\,D_{f_2}(x_2,t)       &      \mbox{\rm ~~~~for $f_2\ne q_i,\bar{q}_i$} \\ 
(N_{i}-1)\,D_{f_2}(x_2,t)      &      \mbox{\rm ~~~~for $f_2=q_i$} \\
(N_{i}+1)\,D_{f_2}(x_2,t)      &     \mbox{\rm ~~~~for $f_2=\bar{q}_i$}\,.~~~~	
\end{array}
\right.
\label{eq:valrule}
\end{eqnarray}
It is important  to emphasize that the momentum and valence quark number sum rules are conserved by the evolution 
equations (\ref{eq:onepdfeq}) and (\ref{eq:twopdfeq}) once they are imposed at an initial value $t_0$. 
If not true, the sum rules will  not be exactly satisfied during evolution.

The sum rules (\ref{eq:momrule}) and (\ref{eq:valrule}) are written with respect to the first parton. 
Assuming the parton exchange symmetry (\ref{eq:partonsym}) to be valid for any value of $t$, 
the sum rules could also be written with respect to
the second parton. In this case,  the integration is performed over $x_2$ up to $(1-x_1)$ with the first parton flavor $f_1$ 
and the momentum fraction $x_1$ fixed. 

\begin{figure*}[t]
\centering\includegraphics[width = 11cm]{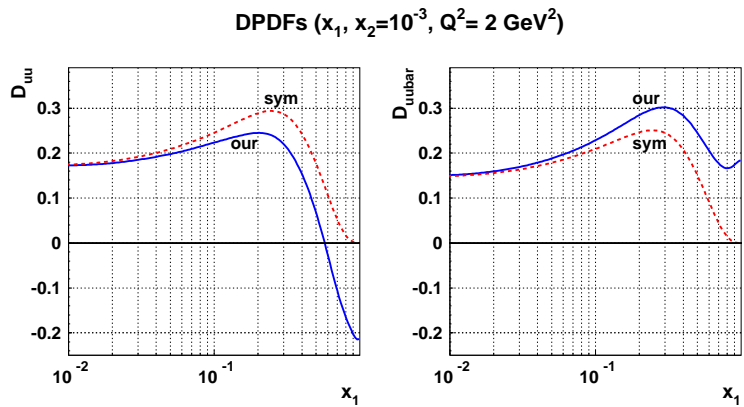}
\caption{The input distributions $D_{uu}$ and $D_{u\bar{u}}$ (multiplied by $x_1x_2$) from   
Eqs.~(\ref{eq:gsinput}) (dashed lines - sym) and (\ref{eq:our3}) (solid lines - our) for fixed $x_2=10^{-3}$.}
\label{fig2}
\end{figure*}

\section{Symmetric initial conditions}

To solve Eqs.~(\ref{eq:onepdfeq}) and (\ref{eq:twopdfeq}) we need to specify initial conditions for
both the SPDFs and DPDFs. The initial SPDFs can be taken from well-established parameterizations, e.g. from
the leading-order (LO) MSTW paramet\-ri\-zation \cite{Martin:2009iq}, which we use in the forthcoming analysis. However,
the specification of the initial DPDFs needs assumptions since the experimental knowledge on the DPDFs is
very limited. For practical reasons, their form
is built out of the existing SPDFs. For example, in Refs.~\cite{Korotkikh:2004bz, Gaunt:2009re} the {\it symmetric}  
initial conditions with respect to the parton interchange (\ref{eq:partonsym}) were proposed,
\begin{eqnarray}\nonumber
\label{eq:gsinput}
 D_{f_1f_2}(x_1,x_2,t_0)\!\!\! &=&\!\!\!
  D_{f_1}(x_1,t_0)\,D_{f_2}(x_2,t_0)
  \\
 &\times&\!\!\!  \frac{(1-x_1-x_2)^2}{(1-x_1)^{2+n_1}(1-x_2)^{2+n_2}},~~~~
\end{eqnarray}
where in the correlation factor, $n_{1,2}=0$ for sea quarks and $n_{1,2}=0.5$ for valence quarks. 
These distributions  are also positive definite provided that the SPDFs are  positive.

\begin{figure*}[t]
\centering\includegraphics[width = 11cm]{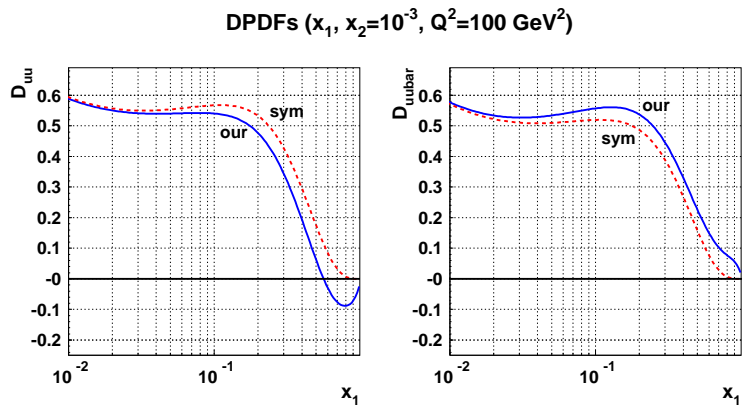}
\caption{The distributions from Fig.~\ref{fig2} evolved up to $Q^2=100~{\rm GeV}^2$.}
\label{fig3}
\end{figure*}

In Fig.\ref{fig1} we show how  ansatz (\ref{eq:gsinput}) fulfills
the momentum and valence quark number sum rules by  plotting the ratio of the rhs to lhs for
Eqs.~(\ref{eq:momrule}) and (\ref{eq:valrule}), respectively. The sum rules are fulfilled if the ratios 
equal one. In the valence number sum rule
$q_i=u$ and $f_2=g,u,\bar{u}$  (for simplicity, we also set  $n_1=n_2=0$). We see that the momentum sum rule is quite well
satisfied while the valence quark number sum rule is significantly violated. 
The limiting values, $3/2$ for $f_2=\bar{u}$ and $1/2$ for $f_2=u$, correspond to the values $(N_u\pm 1)/N_u$ for $N_u=2$, 
which are obtained from Eq.~(\ref{eq:gsinput}) computed for $x_2\ll 1$.  
The case with $n_{1,2}=0.5$ for valence quarks leads to similar results.

\begin{figure*}[t]
\centering\includegraphics[width = 11cm]{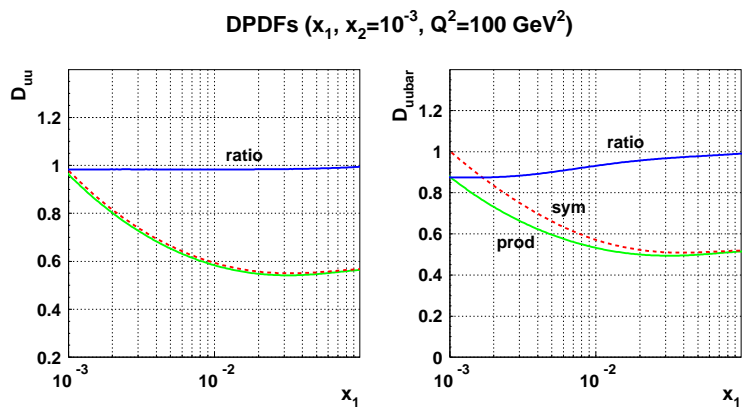}
\caption{The evolved distributions $D_{uu}$ and $D_{u\bar{u}}$ (sym) from the symmetric input (\ref{eq:gsinput})  
compared to the corresponding products of the SPDFs, $D_u$ and $D_{\bar{u}}$ (prod). 
The ratio of the two  distributions is also shown for each case.
}
\label{fig4}
\end{figure*}
\section{Attempt to satisfy sum rules}

Is it possible to construct initial distributions
which {\it exactly} fulfill the discussed sum rules? 
The answer is in the affirmative if we concentrate on the first parton, treating the second one as  a spectator (or vice versa).

To obey the momentum sum rule  (\ref{eq:momrule}), it is enough to  postulate  the following form (we skip
$t_0$ in the notation):
\begin{eqnarray}
\label{eq:our1}
D_{f_1f_2}(x_1,x_2)\,=\, \frac{1}{1-x_2} D_{f_1}\!\!\left(\frac{x_1}{1-x_2}\right)\, D_{f_2}(x_2)\,.
\end{eqnarray}
However, the valence number sum rule (\ref{eq:valrule}) needs corrections for identical quark flavors or antiflavors 
which  do not spoil the already fulfilled momentum sum rule. The form below does the job
\begin{eqnarray}\nonumber
D_{f_if_i}(x_1,x_2)\!\!\!\!\!&=&\!\!\!\!\! \frac{1}{1-x_2} \left\{D_{f_1}\!\!\left(\frac{x_1}{1-x_2}\right) - \frac{1}{2}
\right\} D_{f_1}(x_2)
\\\nonumber
\\\label{eq:our3}
D_{f_i\bar{f}_i}(x_1,x_2)\!\!\!\!\!&=&\!\!\!\!\!\frac{1}{1-x_2} \left\{D_{f_1}\!\!\left(\frac{x_1}{1-x_2}\right) + \frac{1}{2}
\right\} D_{\bar{f}_i}(x_2)
\end{eqnarray}
where $f_i,\bar{f}_i\in\{u,\bar{u},d,\bar{d},s,\bar{s},\ldots\}$ are quark flavor or antiflavors. 
Unfortunately, there is a price to pay. The DPDFs for identical flavors or anti-lavors are not positive definite. For $x_2\ll 1$
\begin{equation}
D_{f_if_i}(x_1,x_2)\approx \left\{D_{f_1}(x_1)-1/2\right\}D_{f_1}(x_2)
\end{equation}
and for all the  existing parametrizations  $D_{f_1}(x_1)< 1/2$ for $x_1$ bigger than some $x_0$. Thus,
$D_{f_if_i}(x_1,x_2)$ is negative in this range.

This is shown
in Fig.~\ref{fig2} where the initial distributions $D_{uu}(x_1,x_2)$
and $D_{u\bar{u}}(x_1,x_2)$ (multiplied by $x_1x_2$) are plotted for fixed $x_2=10^{-3}$. 
They are constructed from the LO MSTW parametrization
of SPDFs at  $Q^2_0=2~{\rm GeV}^2$, using ansatze (\ref{eq:gsinput}) (sym) and (\ref{eq:our3}) (our).
We see that $D_{uu}$ from ansatz (\ref{eq:our3}) is negative for $x_1>0.6$.  Using a numerical program
which we constructed to solve Eqs.~ (\ref{eq:onepdfeq}) and (\ref{eq:twopdfeq}), we show 
in Fig.~\ref{fig3} that this effect does not change when the distributions are evolved up to $Q^2=100~{\rm GeV}^2$.

The proposed form of initial conditions 
is not symmetric with respect to parton interchange  described by Eq.~(\ref{eq:partonsym}). A simple symmetrization,
\begin{eqnarray}
\label{eq:symetric}
D_{f_1f_2}(x_1,x_2) \to D_{f_1f_2}(x_1,x_2)+D_{f_2f_1}(x_2,x_1)\,,
\end{eqnarray}
does not solve the problem since  the discussed sum rules are violated in such a case.  For example, the integration
over $x_1$  as in the sum rule (\ref{eq:momrule}) gives the correct result with the first term in  Eq.~(\ref{eq:symetric}),  
which is spoiled by a nonzero contribution from the second term. Unfortunately, we could not find 
a better symmetrization prescription which conserves the sum rules.

To summarize our attempt, it seems that in  the construction with SPDFs we cannot find the initial distributions 
which fulfill the sum rules in both the variables $x_1$ and $x_2$. Therefore, the  symmetric parametrization
(\ref{eq:gsinput}), discussed at length in ref.~\cite{Gaunt:2009re},  is still the best proposition for applications.

\section{Factorization at small $x$}

The sum rule problem for the initial distribution concerns the large $x$
behavior of the DPDFs.
If parton momentum fractions are small, $x_{1},x_{2} \ll 1$, both parametrizations
of the initial distributions, given by Eqs.~(\ref{eq:gsinput}), (\ref{eq:our1}), and (\ref{eq:our3}), tend to the factorized form
\begin{equation} 
\label{eq:factorized}
D_{f_1f_2}(x_1,x_2,t_0) \approx D_{f_1}(x_1,t_0) D_{f_2}(x_2,t_0)
\end{equation}
where $f_{1,2}$ denote quark flavors, antiflavors, or a gluon. 

Does the approximate factorization hold during the evolution? The  inspection of Eq.~(\ref{eq:twopdfeq})
reveals that in general  the third, splitting term violates  the factorization during the evolution. However,  
the scale of the violation depends on the numerical value of the splitting term in comparison 
to the  values of the first two terms in Eq.~(\ref{eq:twopdfeq}). 

The analysis with our numerical program shows that the violation is significant only for the splitting $g\to q\bar{q}$, due
to a large value of the single gluon distribution $g(x)$ at small $x$. This is illustrated in
Fig.~\ref{fig4}, where $D_{uu}$ and $D_{u\bar{u}}$ evolved to $Q^2=100~{\rm GeV}^2$
from the symmetric input are compared to the corresponding products of  $D_{u}$ and $D_{\bar{u}}$ at the same scale.
The effect of the violation of factorization for small values of $x_{1,2}$ is only seen for $D_{u\bar{u}}$,
while for the distributions, $D_{uu}$ and $D_{ug}, D_{\bar{u}g}, D_{gu}, D_{g\bar{u}}$ (not shown here),
relation (\ref{eq:factorized})  holds very well. The same conclusions are valid for the other quark flavors.
To avoid a possible confusion, let us stress that factorization (\ref{eq:factorized}) 
is not expected to be true at large values of parton momentum fractions (i.e. for $x> 0.1$).

\section{Summary}

The specification of initial conditions for the QCD evolution equations of the DPDFs 
(written in the leading lo\-ga\-rith\-mic approximation)  is not a simple task, 
mainly because of the new sum rules which they should obey. 

In the presented attempt we tried to build the initial DPDFs out of the existing SPDFs treating one of the two partons as a spectator. 
The form which we found obeys the momentum and valence quark number sum rule with respect to the longitudinal momentum fraction  
of the active parton. The difficulty with symmetrization
of the proposed form is the reason why the sum rules are not symmetric with respect to the interchange of partons.
Therefore, the approach \cite{Korotkikh:2004bz, Gaunt:2009re} in which parameters 
the symme\-tric form of initial  DPDFs are optimized to approximately fulfill the sum rules   is still the best one can achieve.

We also discussed the factorization of the DPDFs into a product of two SPDFs at small values of parton
momentum fractions $x_{1,2}$. We showed that  such a factorization is to a good approximation  conserved by 
the  QCD evolution except for the distribution $D_{q\bar{q}}$ for which the splitting contribution $g\to q\bar{q}$  
in the evolution equations (\ref{eq:twopdfeq}) is quite important due to a large gluon distribution $g(x)$ at small $x$. 
For large values of $x_{1,2}$, the factorization is not expected.

\begin{acknowledgments}
This work was supported by the Polish NCN Grants No. DEC-2011/01/B/ST2/03915  and No.  DEC-2012/05/N/ST2/02678 as well as by the Center
for Innovation and Transfer of Natural Sciences and Engineering Knowledge in Rzesz\'ow.

\end{acknowledgments}

\section{REFERENCES}
\bibliographystyle{h-physrev4}
\bibliography{mybib}

\end{document}